\documentclass[11pt]{article}
\usepackage{amsmath,amsfonts,amssymb,amsthm,bm}
\usepackage{mathtools}
\usepackage{geometry}
\usepackage{hyperref}
\usepackage{graphicx}
\usepackage{authblk}
\usepackage[backend=biber,maxbibnames=99]{biblatex} 
\usepackage[normalem]{ulem} 
\usepackage{setspace}
\title{A Dynamic Learning Observatory Reveals the Rapid Salinization of Satkhira, Bangladesh}

\author[1,*]{Showmitra Kumar Sarkar}
\author[2,*]{Sai Ravela}

\affil[1]{Department of Urban and Regional Planning, Khulna University of Engineering and Technology, Khulna-9203, Bangladesh}
\affil[2]{Earth, Atmospheric and Planetary Sciences, Massachusetts Institute of Technology, Cambridge, MA, USA}

\affil[*]{Corresponding authors: Showmitra Kumar Sarkar (mail4dhrubo@gmail.com), Sai Ravela (ravela@mit.edu)}

\date{March 2026}

\begin{document}

\maketitle

\begin{abstract}
Soil salinity is a major environmental challenge in coastal Bangladesh, threatening agricultural productivity and local livelihoods. This study develops a machine-learning-based framework to predict and map soil salinity in Satkhira district by integrating field observations with Landsat-derived spectral indices. A total of 205 soil samples collected during 2024-2025 were used to train an Extreme Gradient Boosting (XGBoost) model, and predictions were further improved using a Generalized Additive Model (GAM). Spatial cross-validation was applied to reduce autocorrelation bias, and bootstrap resampling was used to quantify prediction uncertainty. The results show strong spatial variability of soil salinity, with higher concentrations in the southern and central coastal regions and lower levels in the northern inland areas. Vegetation indices, particularly NDVI, along with salinity-related spectral indicators, were identified as key predictors. 10-year-window peak-exposure maps generated for 2014–2023 reveal recurrent high-salinity zones and a persistent, expanding footprint of moderate-to-high salinity exposure across the central parts of the district. Uncertainty analysis indicates higher variability in coastal zones and improved prediction stability when multi-year datasets are combined. The proposed framework provides a robust and scalable approach for long-term monitoring of soil salinity. It supports climate-resilient agriculture, land-use planning, and evidence-based decision-making in coastal Bangladesh.

\end{abstract}

\section{Introduction}
Soil salinity is an increasingly serious global threat, affecting approximately 33\% of irrigated agricultural lands and 20\% of cultivated areas worldwide \cite{Negacz2022}. The accumulation of salts in the plant root zone significantly diminishes soil productivity and crop yields \cite{Zeorb2019, Umamaheswari2009}. According to projections, salinization could affect more than half of the world's arable land by 2050 if current trends persist \cite{Butcher2016}. Natural events (i.e., storm surges and floods) and human activities (i.e., poor drainage and irrigation systems) contribute to salinization \cite{Jingwei2008}. Additionally, climate change impacts (i.e., sea level rise, altered rainfall patterns, and increased surface evaporation) are responsible for the expansion of salinized lands ~\cite{Uri2018}.

Bangladesh, one of the most climate-vulnerable countries \cite{Sarkar2021}, faces significant challenges with soil salinization, especially in its coastal areas \cite{Sarkar2023a}. The coastal region of Bangladesh contains more than 30\% of the country's arable land, with nearly 53\% of the soils in this area affected by varying levels of salinity \cite{Hasan2019}, posing a considerable risk to national food security. The primary factors contributing to soil salinity in coastal Bangladesh are the lack of freshwater flows from upstream rivers, inconsistent rainfall patterns, tidal amplification, storm surges, and inadequately managed coastal polder systems \cite{Akash2023}. The increasing level of soil salinity is driving the shift from agriculture to shrimp aquaculture and has significant effects on livelihoods and socio-economic conditions \cite{Morshed2021}.

Several soil salinity-related studies in coastal Bangladesh cover a wide range of topics and apply advanced methods and approaches. In terms of detection and modeling, \cite{Sarkar2023b} used machine learning with satellite-based indices, while \cite{Sarkar2023a} applied partial least regression. \cite{Morshed2016} utilized satellite imagery to detect salinity, and later, \cite{Morshed2021} proposed salinity-based land-use zoning. \cite{Kumar2019} analyzed soil salinity and its relation to other soil properties, and \cite{Khan2020} examined salinity levels in coastal areas.

Several studies also link climate change with salinity dynamics. \cite{Fahim2023} investigated climate change-induced salinity intrusion, while (Ashrafuzzaman et al., 2022) explored both current and future salinity intrusion scenarios. \cite{Bhuyan2023} studied the spatio-temporal variability in soil and water salinity across the coastal region. In addition, some research highlights the socio-economic effects of salinity. \cite{Hossain2010, Rezoyana2023} assessed the impact of salinity on local livelihoods, while \cite{Bhuyan2023} examined its broader socio-environmental effects on coastal communities. Regarding adaptation and coping strategies, \cite{Haldar2017}  explored strategies for rice farming, and \cite{Rezoyana2023} discussed broader community responses and salinity impacts in affected areas.

The detection and modeling of soil salinity are critical components of resilience and sustainability in coastal Bangladesh. Previous studies have employed various remote sensing indices with statistical models \cite{Sarkar2023a,Morshed2021} and machine learning techniques \cite{Sarkar2023b}. However, the field sample points used in many of these studies are not well distributed, often concentrated in small areas, outdated, poorly representative of the landscape, and sometimes not compatible with remote sensing imagery. These notable limitations in the quality and distribution of field data may significantly affect the accuracy of soil salinity modeling and prediction.

This study addresses the identified gaps by developing a robust, location-specific, machine-learning-based soil salinity mapping framework that integrates satellite-derived spectral indices with in-situ soil measurements to improve spatial resolution and prediction accuracy. In this research, we collect topsoil samples using a carefully designed sampling strategy, establish a field-based topsoil observatory, and analyze them in the laboratory to determine soil electrical conductivity (EC). We use multiple Landsat-derived spectral indices to represent soil salinity dynamics and employ an Extreme Gradient Boosting (XGBoost) regression model to generate predictive salinity maps. To reduce spatial autocorrelation bias and enhance model robustness, we apply a spatial cross-validation strategy and further improve predictions through post-model calibration using a Generalized Additive Model (GAM). The calibrated framework is applied consistently to generate both recent and historical soil salinity maps, enabling assessment of spatial patterns and salinity exposure over time. The findings of this study are expected to provide valuable insights to help policymakers better understand soil salinity patterns and support evidence-based mitigation and adaptation strategies in coastal Bangladesh.

A central challenge in salinity mapping is the absence of temporally consistent, annually validated observations. Under such conditions, annual maps derived from remote sensing cannot be interpreted as independent realizations of a temporal process. Instead, salinity must be viewed through an exposure-based lens. In this work, we formalize this perspective using a windowed peak-exposure operator that estimates, for each pixel, the maximum salinity experienced within a finite temporal horizon. By adopting an expanding-to-rolling 10-year window, we construct a time-indexed sequence of exposure fields that capture persistence, emergence, and decay of salinity hotspots without requiring year-wise validation. This formulation reframes salinity mapping from state estimation to exposure inference under data constraints.

\section{Methods and Materials}
\subsection{Description of the study area}

\begin{figure}
\centering
\includegraphics[width=\linewidth]{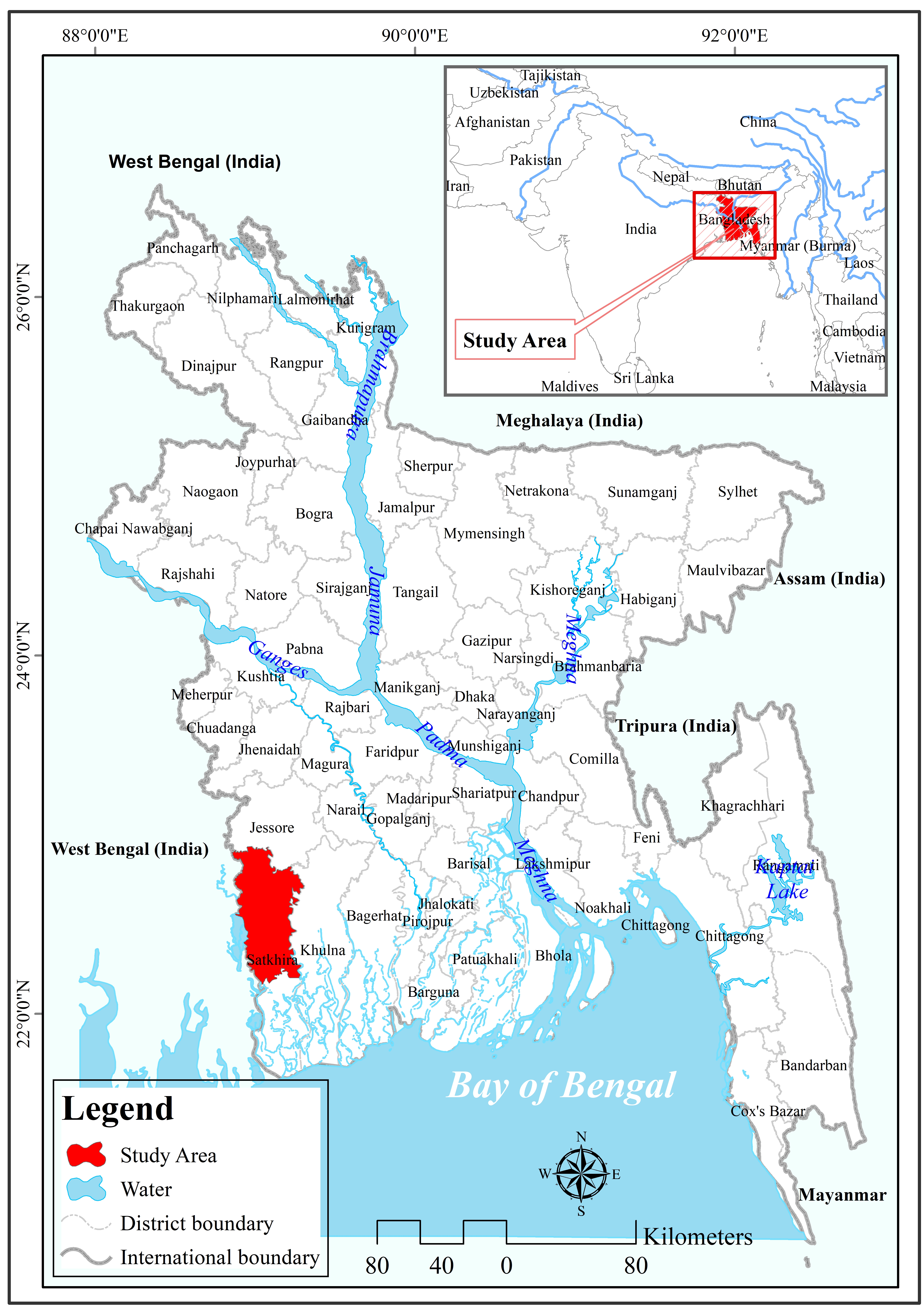}
\caption{Location of the study area}
\label{Figure 1.jpg}
\end{figure}

Satkhira district, located in the southwestern coastal region of Bangladesh, was selected as the study area Figure~\ref{Figure 1.jpg}. The district borders Khulna to the east, Jessore to the north, the Bay of Bengal to the south, and West Bengal, India, to the west. The landscape is predominantly flat and intersected by a dense network of rivers and canals that connect directly to the Bay of Bengal, making the region highly susceptible to tidal influence and salinity intrusion.

Satkhira is widely known for shrimp aquaculture and contributes significantly to national saltwater shrimp production. The district frequently experiences cyclones, storm surges, coastal flooding, heavy rainfall, and heat waves, which exacerbate environmental stress and salinity conditions. Soil salinity is a major environmental challenge in this region and has substantial implications for agricultural productivity and land-use dynamics.
The district has a population of approximately 2.2 million \cite{BBS2024} and is administratively divided into seven upazilas. Local livelihoods are largely dependent on agriculture and shrimp farming. Increasing soil salinity has accelerated the conversion of agricultural land to shrimp aquaculture due to declining crop yields, thereby affecting food security and socio-economic conditions.

We selected Satkhira due to its environmental diversity and high vulnerability to climate change-induced hazards, particularly salinity intrusion and coastal flooding. Focusing on a district-level analysis allows for the development of location-specific insights and policy recommendations for sustainable land and water management. The Sundarbans mangrove forest was excluded from the analysis because of its distinct ecological characteristics and land-cover conditions.

\subsection{Description of the field samples}
We use two types of data in this study: field data and remote sensing data to predict soil salinity. We collected 86 soil samples in March 2024 \cite{Sarkar2024Dataset} and 119 in April 2025 \cite{Sarkar2025Dataset} from different parts of the district, following the protocol described in \cite{sarkar2025topsoil}. We select sampling sites that are free from obstacles such as trees and buildings and are mainly located in dry, open, and fallow fields. We collect topsoil samples from the root zone at a depth of 0–30 cm. We record the geographic locations of all sampling points using Google My Maps. We analyze all soil samples at the ESSG-WECG Laboratory at Khulna University of Engineering \& Technology. We follow the standard operating procedure for measuring soil electrical conductivity using a soil–water ratio of 1:5, as recommended by the Food and Agriculture Organization of the United Nations \cite{FAO2021}. We measure electrical conductivity (EC) using a Hanna HI 6321 benchtop conductivity meter.

\subsection{Description of the remote sensing indices} 
We use Landsat 8 Operational Land Imager (OLI) and Thermal Infrared Sensor (TIRS) Surface Reflectance data to derive spectral indices related to soil salinity. We process all satellite data using the Google Earth Engine (GEE) platform. We acquire Landsat 8 Collection 2, Tier 1 Level-2 Surface Reflectance images for the dry season (1 February to 30 April each year) from 2014 to 2025. We filter images using World Reference System-2 (WRS-2) Path 138 and Row 44, and limit cloud cover to less than 15\%. We clip all images to the Area of Interest (AOI). We use surface reflectance bands, including blue, green, red, near-infrared (NIR), shortwave infrared-1 (SWIR1), shortwave infrared-2 (SWIR2), and thermal infrared, for analysis. We develop 13 different indices (i.e., Salinity Index 1 (SI-1), Salinity Index 2 (SI-2), Salinity Index 3 (SI-3), Salinity Index 11 (SI-11), Intensity Index 1 (INT-1), Intensity Index 2 (INT-2), Brightness Index (BI), Soil Adjusted Vegetation Index (SAVI), Normalized Difference Vegetation Index (NDVI), Enhanced Vegetation Index (EVI), Spectral Ratio Index (RATIO), Blue band reflectance (BLUE), and Near Infrared reflectance (NIR)) applied previously by \cite{Sarkar2023b}. We generate all bands and indices at a spatial resolution of 30 m. We export all outputs as GeoTIFF files for further analysis and modeling.

\subsection {Modeling framework of soil salinity mapping for 2024 and 2025}
\begin{figure}
\centering
\includegraphics[width=\linewidth]{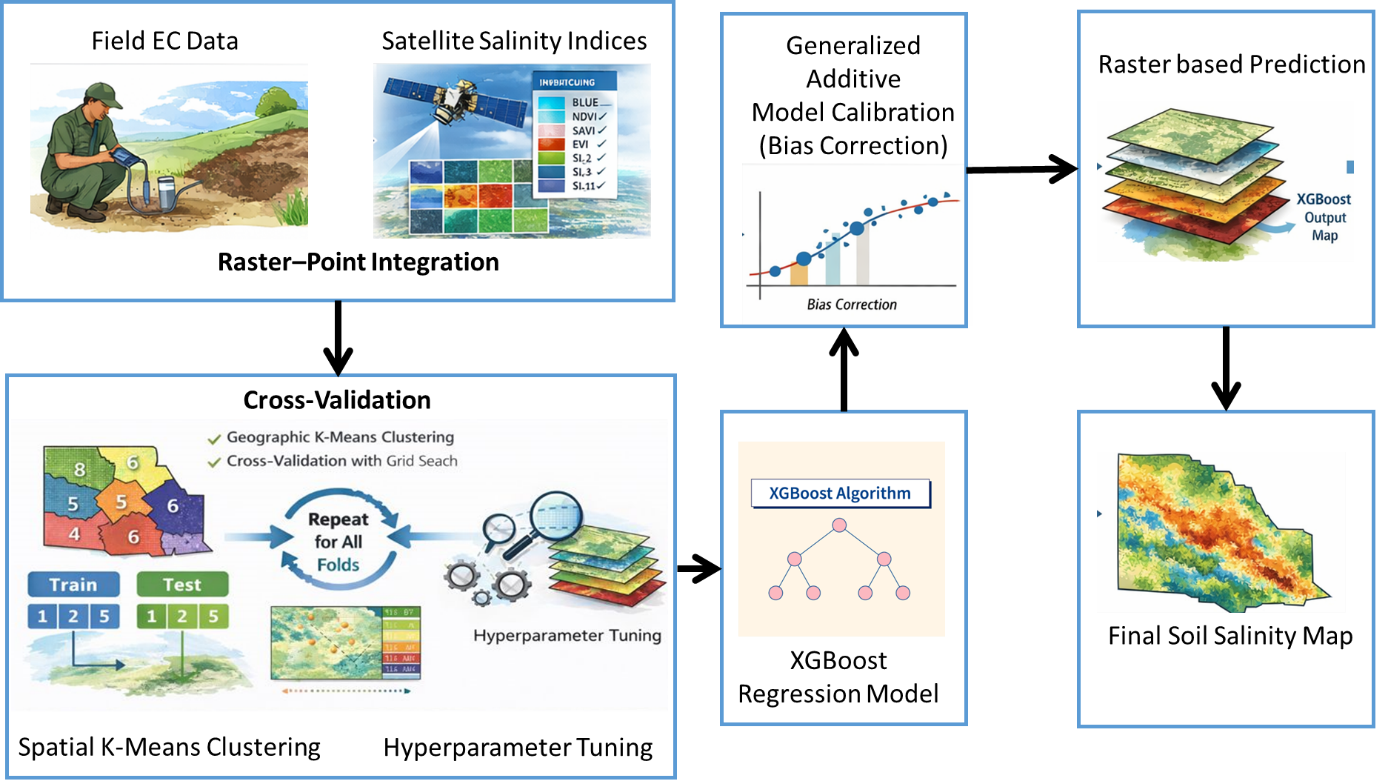}
\caption{Modeling framework of soil salinity mapping}
\label{Figure 2.png}
\end{figure}

The overall modeling framework for soil salinity mapping is illustrated in Figure~\ref{Figure 2.png}. We develop two-year-specific predictive models for 2024 and 2025 using corresponding field observations and satellite-derived predictor variables.

\subsubsection{Dataset construction}
We developed a predictive modeling framework by integrating field-measured soil electrical conductivity (EC) data with Landsat-derived spectral indices. Field-measured EC values (mS/cm) were used as the response variable. Thirteen spectral indices were used as predictor variables. We extracted raster values of all spectral indices at soil sampling locations and merged them with the corresponding EC observations. Samples containing missing predictor values were removed to ensure consistency and reliability of the modeling dataset.

\subsubsection{Spatial cross-validation strategy}
To minimize the effects of spatial autocorrelation and avoid overoptimistic estimates of model performance, we implemented a spatial cross-validation strategy. Geographic coordinates of soil sampling locations were clustered into six spatial groups using the K-means clustering algorithm (k = 6, nstart = 50). A fixed random seed (seed = 123) was used to ensure reproducibility. Each cluster was treated as an independent spatial fold. In each cross-validation iteration, five clusters were used for model training and one cluster for validation. This process was repeated until all clusters were used once as validation data. Model training and validation were controlled using a cross-validation framework that stored all predictions and enabled fold-wise performance evaluation.

\subsubsection{XGBoost model development}

We employed Extreme Gradient Boosting (XGBoost) regression to model soil salinity because it captures nonlinear relationships and interactions among predictor variables. Model implementation was carried out using the caret framework. Model performance was evaluated using Root Mean Squared Error (RMSE).

Hyperparameter tuning was conducted using a grid-search approach. The following parameter ranges were evaluated:

\begin{itemize}
    \item Number of boosting rounds (nrounds): 500, 1000, 1500
    \item Maximum tree depth (max\_depth): 3, 5, 7
    \item Learning rate (eta): 0.01, 0.005, 0.001
    \item Minimum child weight (min\_child\_weight): 5, 10, 15
    \item Gamma (minimum loss reduction): 0
    \item Column subsampling rate (colsample\_bytree): 1.0
    \item Row subsampling rate (subsample): 1.0
\end{itemize}

The optimal model was selected based on the lowest RMSE across spatial validation folds. The final XGBoost model was then used to generate soil salinity predictions at the sampling locations.

\subsubsection{Post-model calibration Using GAM}
To reduce systematic bias in the XGBoost predictions, we applied a post-model calibration step using a Generalized Additive Model (GAM). To avoid optimistic bias, the GAM was fitted using out-of-fold predictions obtained during spatial cross-validation. Specifically, observed EC values were regressed against cross-validated XGBoost predictions using a smooth spline function to capture nonlinear relationships.
The calibrated relationship was then applied to full-model predictions for spatial mapping. This approach prevents information leakage during calibration and ensures that model evaluation metrics reflect predictive performance on spatially independent data.

\subsubsection{Spatial prediction and mapping}
We generated spatially continuous soil salinity maps by applying the optimized XGBoost model to a raster stack of all predictor indices. The GAM calibration function was then applied pixel-wise to the predicted raster to obtain calibrated soil salinity estimates. To ensure spatial consistency, the Area of Interest (AOI) was projected to match the raster coordinate reference system. The final calibrated soil salinity map was cropped and masked using the AOI boundary. All outputs were produced at a spatial resolution of 30 m.

\subsection{Bootstrap-based uncertainty assessment} 

To quantify prediction uncertainty associated with the machine learning model, we applied a bootstrap-based resampling approach. Bootstrap resampling allows assessment of model sensitivity to variations in the training data and provides an empirical estimate of prediction uncertainty without strong distributional assumptions. We generated 100 bootstrap realizations by randomly resampling the training dataset with replacement. For each bootstrap iteration, a new Extreme Gradient Boosting (XGBoost) model was trained using the same optimal hyperparameters as the final calibrated model. This ensured that uncertainty estimates reflect variability arising from the data rather than changes in model structure or parameterization.

Each bootstrap-trained XGBoost model was applied to the full stack of Landsat-derived spectral indices to produce a spatial prediction of soil salinity. The same post-model calibration function based on the Generalized Additive Model (GAM) was applied to all bootstrap predictions to maintain consistency with the final salinity maps. The ensemble of calibrated bootstrap prediction maps was combined into a raster stack, and pixel-wise standard deviation was calculated across all bootstrap realizations. This standard deviation represents spatial prediction uncertainty, where higher values indicate greater variability among bootstrap predictions and lower confidence in model estimates. The resulting uncertainty map was cropped and masked to the Area of Interest to ensure spatial consistency with the predicted soil salinity maps. This bootstrap-based uncertainty assessment provides a spatially explicit representation of model uncertainty and complements residual-based evaluation by highlighting regions where predictions are more sensitive to sampling variability.

An uncertainty assessment was conducted across multiple sampling scenarios to evaluate the influence of training data availability. Separate analyses were performed using (i) 2024 samples only, (ii) 2025 samples only, and (iii) a combined dataset of 2024 and 2025 samples. The comparison of these scenarios demonstrated that uncertainty decreased when the combined dataset was used, indicating improved model stability and prediction reliability. Therefore, the combined sample dataset was adopted for decadal soil salinity mapping. 

\subsection{Decadal Soil Salinity Mapping and Peak-Exposure Analysis (2014-2023)} 

To generate soil salinity maps for the historical period from 2014 to 2023, we applied the same modeling framework used for the 2024 and 2025 predictions. We used a combined soil salinity dataset that integrates all available field observations to represent the spatial variability of soil salinity across the study area. This combined dataset was used to develop a robust model structure that was then applied uniformly across the decadal period.

For each year between 2014 and 2023, we prepared year-specific Landsat-derived spectral indices using the same preprocessing and index-calculation procedures. These year-specific satellite datasets were used as model inputs, with the trained model's structure and parameters left unchanged. This approach ensures methodological consistency across years; however, the resulting maps represent model-applied estimates rather than independently validated annual salinity states.

We applied the trained model to the annual spectral index raster stacks to generate year-wise soil salinity predictions. The post-model calibration step was then applied to each yearly prediction using the same calibration function. All predicted maps were produced at a spatial resolution of 30 m and clipped to the Area of Interest. This consistent modeling strategy enables reliable assessment of spatial patterns and decadal salinity exposure across the study period.

Because in situ EC measurements were available only for 2024--2025, year-specific salinity maps prior to 2024 cannot be interpreted as independently validated annual states. Under this constraint, we reformulate salinity mapping as an exposure inference problem. Specifically, we define a peak-exposure operator over an expanding-to-rolling 10-year window, in which the value at each pixel and year $t$ is the maximum predicted salinity over the preceding decade (or, initially, over fewer years). This construction yields a time-indexed sequence of exposure fields that capture the spatial footprint of extreme salinity within a finite horizon, rather than attempting to reconstruct annual states. This metric represents the highest inferred dry-season salinity exposure within a decadal horizon and constitutes an exposure envelope rather than a reconstruction of annual salinity dynamics.

The expanding-to-rolling peak-exposure formulation provides a distinct interpretation of salinity dynamics under observational constraints. Rather than attempting to reconstruct annual salinity states—which are not independently validated—the method captures the spatial footprint of extreme exposure over a finite horizon. This identifies areas experiencing persistent or recurring high salinity, as well as regions where exposure declines when previously observed extreme values fall outside the rolling window. In this sense, the framework reveals not only the presence of salinity hotspots, but their persistence and turnover, providing a basis for risk assessment that does not rely on independently validated annual salinity fields.

Formally, this formulation can be expressed as follows. Let $S(x,t)$ denote the model-predicted salinity at location $x$ and year $t$. The peak-exposure field is defined as:

\begin{equation}
E(x,t) = \max_{t' \in [\max(t-9, t_0),\, t]} S(x,t')
\end{equation}

where $t_0 = 2014$ denotes the initial year of the record. This operator induces an expanding window for $t < t_0 + 9$ and a rolling window thereafter. The resulting field $E(x,t)$ represents the maximum inferred salinity exposure within the active window.

\section{Results}

\subsection{Spatial distribution of field samples and remote sensing indices }

\begin{figure}
\centering
\includegraphics[width=\linewidth]{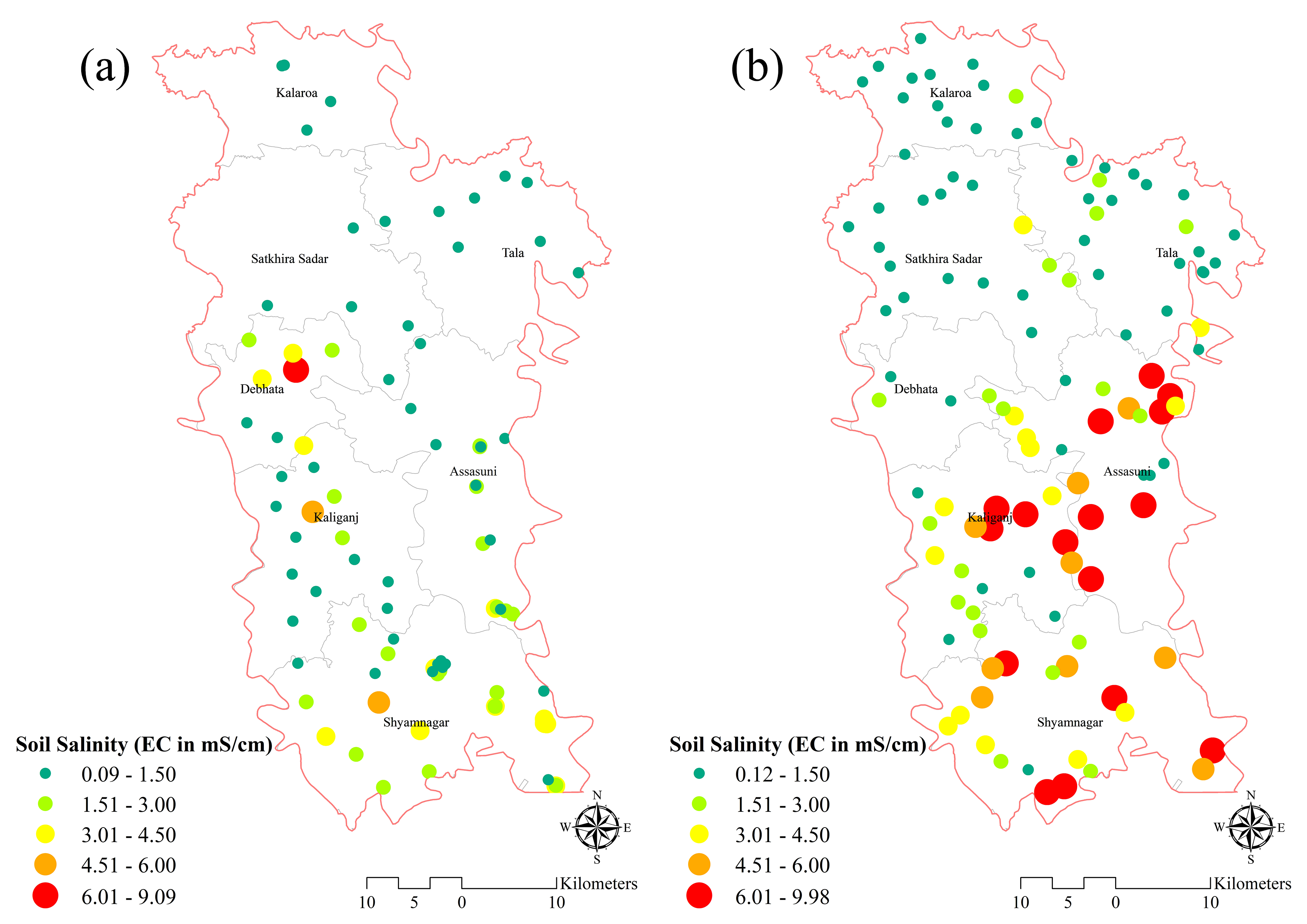}
\caption{Spatial distribution of soil salinity (EC, mS/cm) samples in Satkhira District}
\label{Figure 3.png}
\end{figure}

The bubble density maps Figure~\ref{Figure 3.png}show the spatial distribution of soil electrical conductivity (EC, mS/cm) in 2024 and 2025. In 2024, soil salinity ranged from 0.09 to 9.09 mS/cm and followed a clear north-south pattern. The northern and northwestern parts of Satkhira, including Kalaroa and Satkhira Sadar, mostly showed EC values between 0.09 and 1.50 mS/cm. Values ranging from 1.51 to 3.00 mS/cm were observed in the central part of the district and in parts of Assasuni and Kaliganj. EC values of 3.01-4.50 mS/cm and 4.51-6.00 mS/cm were mainly observed around Debhata, Kaliganj, Assasuni, and Shyamnagar. The highest EC values (6.01-9.09 mS/cm) were mainly concentrated near Debhata.

In 2025, soil salinity ranged from 0.12 to 9.98 mS/cm, with a wider range of higher EC values than in 2024. The northern part of Satkhira still showed lower EC values, mostly between 0.12 and 2.00 mS/cm. Values from 2.01 to 4.00 mS/cm were widely distributed across central and eastern areas, including Tala and Debhata. EC values of 4.01-6.00 mS/cm and 6.01-8.00 mS/cm were common in Assasuni, Kaliganj, and Shyamnagar. The highest EC values (8.01-9.98 mS/cm) were found in several locations in Kaliganj, Assasuni, and coastal Shyamnagar. Overall, the 2025 map shows that higher EC values are more widespread across the central and southern parts of Satkhira than in 2024, while the northern areas remain relatively less affected.

Figure~\ref{Figure 4} shows how 13 soil salinity-related indices were spread out across the study area in 2024 and 2025. Three of these indices (Salinity Index 1 (SI 1), Salinity Index 3 (SI 3), and Band 2) exhibit a fairly stable spatial pattern across both years. Most of the lower values of these indices are found in the northern part of Satkhira, while most of the higher values are found in the southern coastal area near the Sundarbans and in some central areas. This difference between the north and south remains clear in 2024 and 2025, but in 2025, the southern region appears to have higher values across a larger area.

The other indices, on the other hand, show more spatial variability across the study area in both years. This is because the surface characteristics are different. The Salinity Index 11 (SI 11), Soil Adjusted Vegetation Index (SAVI), Near Infrared (NIR), Normalized Difference Vegetation Index (NDVI), and the RATIO index all show the same pattern in space. The northern part of the district has higher values, while the central and southeastern parts have lower values. Overall, the spatial patterns of these indices for both 2024 and 2025 show clear regional differences across Satkhira, particularly between the northern inland areas and the southern coastal belt.

\subsection{Predictor importance and model validation}

We used the model to assess the importance of the top 10 criteria for 2024 and 2025. Figure~\ref{Figure 5}. NDVI has the maximum significance value of 0.753 in 2024 Figure~\ref{Figure 5}. This suggests that it is the most essential factor for forecasting soil salinity. The primary contribution comes from the BLUE band (0.092), SI 11 (0.055), and INT 1 (0.050). Other variables, such as BI (0.017), EVI (0.012), and NIR (0.008), are less crucial. INT 2 and SI 2 don't do much, while SI 3 doesn't do anything at all.
A similar trend may be observed in 2025 (Figure 5(b)), with NDVI remaining the most significant predictor (0.721). The next most essential factors are INT 1 (0.084) and SI 11 (0.064).

The model showed strong explanatory power when tested with spatial cross-validation, achieving $R^2$ values of 0.912 in 2024 and 0.856 in 2025. The scatter plots of actual vs. expected values reveal that most points lie near the line of perfect prediction (y = x). This means that the predicted and observed values are in excellent agreement. The model achieved a coefficient of determination ($R^2$) of 0.912 for 2024 (Figure~\ref{Figure 6}), indicating that it explains more than 91\% of the variation in soil salinity. The model also had a low root mean squared error (RMSE) of 0.454, indicating it was quite accurate in its predictions. The model achieved an $R^2$ of 0.856 for 2025 Figure~\ref{Figure 6}, indicating that it explains around 86\% of the variation in soil salinity. The RMSE of 0.997 indicates that the predictions are slightly worse than in 2024 but remain very good. The validation findings for both years show that the model makes stable and consistent predictions when soil salinity is cross-validated throughout the research region. The forecasts for 2024 were marginally more accurate than those for 2025. Prediction errors increase slightly as salt levels rise, indicating that the model is less accurate in difficult situations.

The residual plots for 2024 and 2025 show how prediction errors are distributed across the predicted values. Most of the residuals in 2024 Figure~\ref{Figure 7} are very close to the zero reference line, indicating that the observed and predicted soil salinity values are very similar. The residuals are randomly distributed and not very large, which means the model is stable and has little systematic bias.

In 2025, Figure~\ref{Figure 7}, the residuals remain mostly around the zero line but are slightly more spread out, especially at higher predicted EC values. This shows that the predictions are slightly more variable at higher salinity levels than in 2024. The fact that the residuals are randomly distributed in both years, on the other hand, suggests that the model performs consistently without making large errors in either direction. The residual patterns show that the model can be trusted to predict soil salinity for both 2024 and 2025.

\subsection{Spatial prediction of soil salinity}

Figure~\ref{Figure 8} and Figure~\ref{Figure 9} show the distribution of modeled soil salinity in the Satkhira district in 2024 and 2025, respectively. The salinity levels will range from 0.08 to 7.19 mS/cm in 2024. Salinity levels are higher in the southern and southwestern parts of the district, especially in Shyamnagar, Assasuni, and some parts of Kaliganj and Debhata. The northern upazilas, such as Kalaroa and Tala, have lower salinity levels than the central areas of Satkhira Sadar and Debhata, which have mixed salinity levels.

The highest salinity level in 2025 is higher than in 2024, with values ranging from 0.00 to 9.86 mS/cm. Higher salinity levels are more widespread in the southern and southeastern parts of the country, especially in Shyamnagar, Assasuni, and Kaliganj. Some central areas also have higher salinity levels than they did last year. The northern part of the district still has lower salinity levels.

They used the same classification scheme (<1.5, 1.5–3, and >3 mS/cm) to compare the two maps. The difference between 2024 and 2025 is not due to differences in classification, but to differences in modeled salinity patterns. It is important to remember that the model is based on soil observations and is meant to show how salty the land is. It doesn't make sense to make predictions over open bodies of water, like rivers and canals, and you shouldn't take them to mean that the water is really salty. The spatial patterns for both years show a clear salinity gradient from north to south. The southern coastal belt has higher salinity, while the northern inland areas have lower salinity.

\subsection{Uncertainty analysis of soil salinity prediction}

Figure~\ref{Figure 10} shows how the model's uncertainty (standard deviation) is distributed across space for soil salinity estimates derived from field samples collected in 2024, 2025, and the combined dataset. The central and southern regions of Satkhira, such as Kaliganj, Assasuni, and Shyamnagar, are where the most uncertainty (>1) occurs in 2024. Most of the sites with the least amount of uncertainty (<0.5) are in the north. When you put the samples from 2024 and 2025 together, the regions with a lot of uncertainty become smaller in certain places. This suggests that the model is more stable with more data.

The southern and southeastern coastal areas are still less certain than the northern upazilas in 2025. In much of the study area, particularly in the middle regions, the combined data for 2024 and 2025 make the patterns much less obvious. The maps show that, in general, increasing the sample size and time period improves model accuracy and reduces forecast uncertainty throughout the district.

\subsection{Decadal soil salinity exposure mapping (2014-2023)}

The maps generated for the period 2014-2023 (see Figure~\ref{Figure 11}) show how the highest salinity levels in the Satkhira district are expected to evolve over the next ten years. Throughout the study period, there is a consistent north-south gradient. Kalaroa and Tala are two of the northern upazilas that typically have lower salinity (less than 1.5 mS/cm). The middle and southern parts of the region, on the other hand, contain more salt. 

Debhata, Assasuni, and portions of Kaliganj are in the central and southwestern areas of the nation, where there are the most sites with high salinity (greater than 3 mS/cm). Shyamnagar and Assasuni are two places in the southeast and on the coast where there are a number of these regions. The salt levels in the core sections remain variable, ranging from moderate to high. There is still a lot of salt in the southern and southeastern upazilas, and clear salt hotspots remain in Shyamnagar, Assasuni, and Kaliganj. 

The northern part of the district always has less salt than the southern coastal areas. The maps from 2014 to 2023 show that the southern coastal portion of Satkhira has consistently had the highest salt exposure, while the northern interior areas have had lower exposure. These patterns show that the distribution of soil salinity in the area investigated is greatly affected by how near the land is to the sea, how tides work, and how the land is used.

The 10-year window peak-exposure maps quantify the spatial extent of salinity-affected regions across the district, see Figure~\ref{Figure 12}. From 2014 to 2023, salt levels in the study area were between 1.5 and 3 mS/cm at least once in around 22.02\% of the area. About 38.13\% had values exceeding 3 mS/cm, suggesting that salinity levels were typically moderate to high. Most of the high-exposure locations are in Assasuni, Debhata, Shyamnagar, and Kaliganj. These areas account for approximately 88.79\% of the overall high-salinity zone. These data illustrate how much more area peak salinity exposure occupies than is displayed in the state each year. Figure 11 illustrates how the area (\%) is divided up into three salinity groups for each upazila: less than 1.5 mS/cm, 1.5–3 mS/cm, and more than 3 mS/cm. This is based on the highest level of exposure observed at least once during 2014-2023.

\section{Discussion}

Using field observations and Landsat-derived spectral predictors, we created a machine-learning-based framework to predict and map soil salinity in the coastal district of Satkhira. The findings indicate that soil salinity exhibits substantial spatial variability, with concentrations consistently greater in southern and central coastal regions and substantially lower in northern inland areas. These spatial patterns are closely linked to tidal intrusion, proximity to saline rivers, aquaculture expansion, and coastal hydrodynamic processes that facilitate salt accumulation in surface soils \cite{Akash2023,Morshed2016}.

The spatial distribution of spectral indices indicates the distinct environmental gradients present throughout the study area. In southern littoral zones, salinity-related indices demonstrate higher values, while vegetation indices exhibit lower values in these regions as a result of salinity-induced vegetation stress. The modeling results underscore the importance of vegetation-related predictors, particularly NDVI, EVI, and SAVI, highlighting that vegetation condition is a critical factor in explaining variability in soil salinity. The fact that NDVI is the most significant predictor in the predictor importance analysis implies that the vegetation condition is a reliable proxy for salinity stress in the study area. NDVI values are reduced in coastal agricultural systems due to elevated soil salinity, which in turn reduces vegetation vigor. Therefore, it is probable that NDVI captures the ecological manifestation of salinity rather than merely detecting soil reflectance signals. Although this improves predictive performance, it also suggests that the model partially emulates vegetation-salinity interactions and land-use responses, particularly in regions transitioning to aquaculture and other land-use changes. This also suggests that the model may not accurately depict soil salinity in areas undergoing land-use transitions, as it may not be able to distinguish between soil salinity and vegetation conditions.

Additionally, salinity indicators, including SI-11 and intensity indices, made substantial contributions by capturing soil reflectance and moisture characteristics associated with saline environments. These results are consistent with prior research conducted in coastal Bangladesh, which has documented robust correlations between soil salinity conditions and spectral indicators \cite{Morshed2021,Sarkar2023a}.

XGBoost captured the nonlinear interactions between spectral predictors and soil salinity within a GAM-calibrated framework, which exhibited strong explanatory power in spatial cross-validation. This confirms the efficacy of machine learning techniques. The spatial prediction maps revealed persistent salinity hotspots in southern upazilas, particularly Shyamnagar, Assasuni, and Debhata. These hotspots are significantly impacted by tidal inundation, saline water intrusion, and aquaculture activities. The vulnerability of these coastal landscapes to salinity intrusion has been further emphasized by the documentation of comparable spatial patterns in previous studies \cite{Morshed2021,Sarkar2023a,Sarkar2023b}. The model's predictive performance may fluctuate in unseen regions or under varying environmental conditions, despite its strong performance during spatial validation.

The analysis of the time-indexed 10-year window emphasizes persistent spatial patterns and regions of recurrent high-salinity exposure between 2014 and 2023, rather than salinity dynamics specific to a single year. Moderate-salinity areas progressively emerged in the district's central regions, while high-salinity zones remained concentrated in the southern coastal regions. This persistence reflects sustained exposure to high salinity within the windowed framework, rather than direct observation of recovery dynamics, which are subject to tidal and climatic influences, leading to the accumulation and discharge of salt over several years \cite{Bhuyan2023}. The observed windowed persistence is consistent with field observations that suggest that soil salinity in coastal agricultural systems rarely decreases abruptly.

Uncertainty assessment provides additional insights into the model's reliability. The southern and southeastern coastal regions exhibit greater prediction variability, as salinity dynamics are complex and influenced by a range of environmental factors. The increased uncertainty in these regions is likely due to greater heterogeneity in hydrological and land-use conditions, including tidal influence, aquaculture ponds, and micro-topographic variability. These factors introduce localized and dynamic salinity patterns that are more challenging for the model to capture consistently. Conversely, the reduced uncertainty in northern inland regions suggests that the environmental conditions are more stable and the salinity patterns are more predictable. The significance of temporally distributed sampling in enhancing model stability and prediction accuracy is underscored by the observed reduction in uncertainty when integrating multi-year datasets. These results underscore the necessity of adaptive sampling strategies and continuous monitoring in salinity-prone regions.

Unlike previous salinity studies in coastal Bangladesh that rely on single-year modeling or limited field observations, this study integrates a structured field-based salinity observatory, spatial cross-validation, post-model calibration, uncertainty quantification, and decadal mapping within a unified framework. This comprehensive method enhances the reliability of predictions and facilitates the long-term monitoring of salinity dynamics in coastal landscapes.
The results have significant implications for the sustainable management of land and resources. Targeted interventions, including the adoption of salt-tolerant crops, enhanced irrigation management, and land-use planning in vulnerable coastal zones, can be facilitated by identifying salinity hotspots. The modeling framework can also support climate-resilient agricultural planning and early warning systems for salinity intrusion amid evolving climatic and hydrological conditions \cite{Haldar2017,Rezoyana2023}.

However, it is imperative to recognize that these contributions are not without their limitations. Although field sampling was spatially structured, the number of samples may still be insufficient to capture fine-scale salinity variability across heterogeneous landscapes. Particularly in dispersed agricultural and aquaculture regions, the detection of microscale salinity variations may be limited by the use of medium-resolution satellite imagery (30 m). The analysis predominantly reflects dry-season conditions. We did not model the influences of rainfall, river discharge, and the tidal cycle on seasonal salinity. 
Furthermore, environmental variables, including groundwater salinity, precipitation, temperature, land use, and proximity to saline water sources, were not considered, which may have restricted the explanatory capacity. Consistent modeling assumptions and spectral predictors were also employed in historical mapping, which may have introduced uncertainty in the representation of long-term dynamics under changing climatic and land-use conditions.

This study demonstrates the efficacy of combining field observations, satellite-derived spectral indices, and advanced machine learning techniques for monitoring soil salinity in climate-vulnerable coastal environments. The proposed method offers a framework that is both transferable and scalable, designed to facilitate evidence-based land management, agricultural planning, and policy interventions in coastal Bangladesh and similar deltaic regions. This framework maps salinity patterns. It is crucial to interpret the results within a hazard-envelope framework that depicts peak salinity exposure over time, rather than as a reconstruction of annual salinity states.

\section{Conclusion}

By combining field observations with Landsat-derived spectral predictors, this study developed a machine-learning-based framework to predict and map soil salinity in the Satkhira district of coastal Bangladesh. The XGBoost model, along with post-model calibration using a Generalized Additive Model (GAM), was very good at capturing the spatial variability and nonlinear relationships that come with soil salinity. The framework identified areas of high salinity, particularly in the southern and central coastal regions, which shows how tidal intrusion, the growth of aquaculture, and coastal hydrological processes affect salinity levels. 
The creation of time-indexed 10-year window peak-exposure maps (2014–2023) showed that salinity exposure remains spatially concentrated over time and that salinity-affected areas are repeatedly exposed, indicating persistent spatial concentration. The uncertainty assessment made even clearer where predictions are more sensitive to sampling variability. It showed that combining field observations from multiple years makes models more stable and predictions more reliable.

The combination of a structured field-based salinity observatory, satellite-derived spectral indices, spatial cross-validation, and uncertainty analysis creates a robust and flexible system for monitoring soil salinity over time. The results have significant ramifications for climate-resilient agriculture, sustainable land-use planning, and resource management in coastal Bangladesh. Even with these contributions, some problems still exist. It wasn't clear how seasonal changes in salinity were modeled, and environmental factors such as groundwater salinity, climate change, and land-use changes weren't fully accounted for. Future studies that combine observations from multiple seasons, hydroclimatic data, and high-resolution remote sensing could improve our ability to make predictions and better understand processes.
The suggested framework is a useful and adaptable way to map and monitor changes in soil salinity in coastal areas vulnerable to climate change. It also helps people make evidence-based decisions for sustainable development and adaptation planning.

\section*{Data Availability}
Zenodo hosts the salinity data from 2024~\cite{Sarkar2024Dataset} and 2025~\cite{Sarkar2025Dataset}. 

\section*{Acknowledgment}

This research is part of the MIT Climate Grand Challenges, Jameel Observatory CREWSNet, and Weather and Climate Extremes projects. Funding from Schmidt Sciences, LLC, and Bangkok Bank is gratefully acknowledged. The authors used OpenAI's ChatGPT as a writing assistant for language polishing and for suggesting edits to the exposition. The authors also used Grammarly for editing. All technical content, claims, and errors remain the responsibility of the authors.

\nocite{*}
\printbibliography

\newpage

\begin{figure}[p]
\centering
\includegraphics[width=5truein]{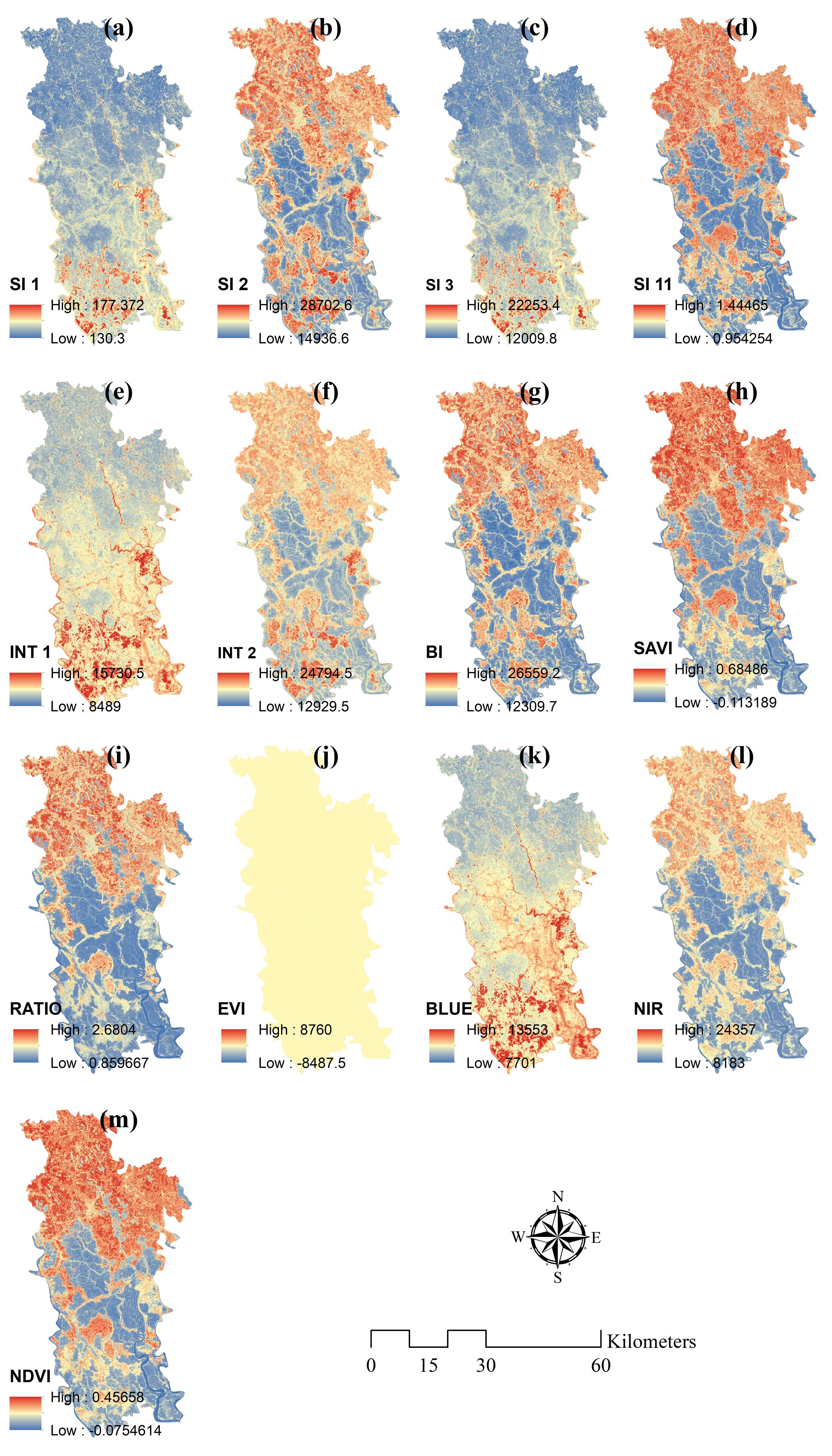}
\caption{Spatial distribution of Landsat-derived soil salinity indices across the study area for 2024 (a-m) and 2025 (n-z)}
\label{Figure 4}
\end{figure}

\clearpage

\addtocounter{figure}{-1}

\begin{figure}[p]
\centering
\includegraphics[width=5truein]{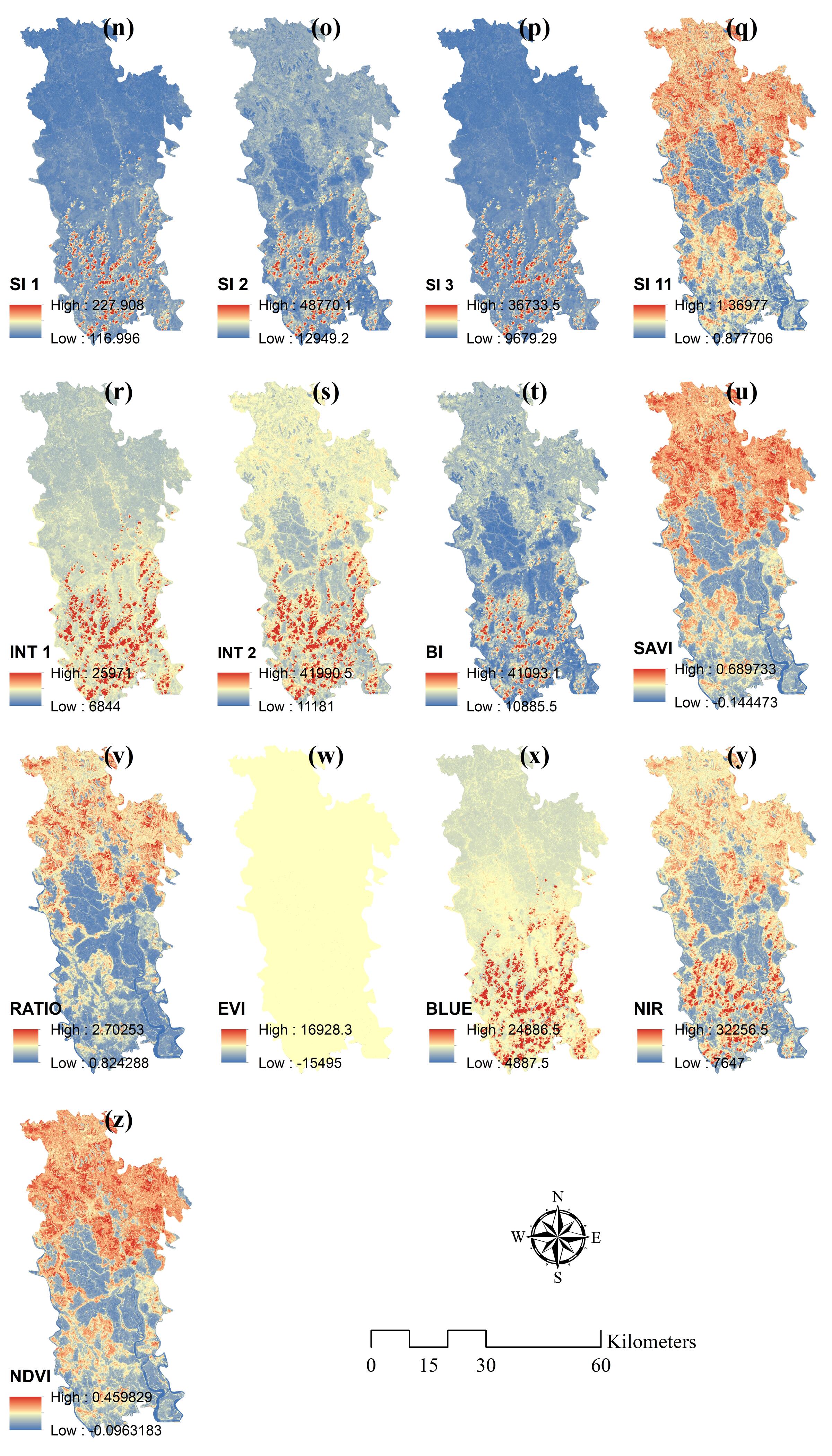}
\caption{(continued)}
\label{Figure 4}
\end{figure}

\begin{figure}
\centering
\includegraphics[width=\linewidth]{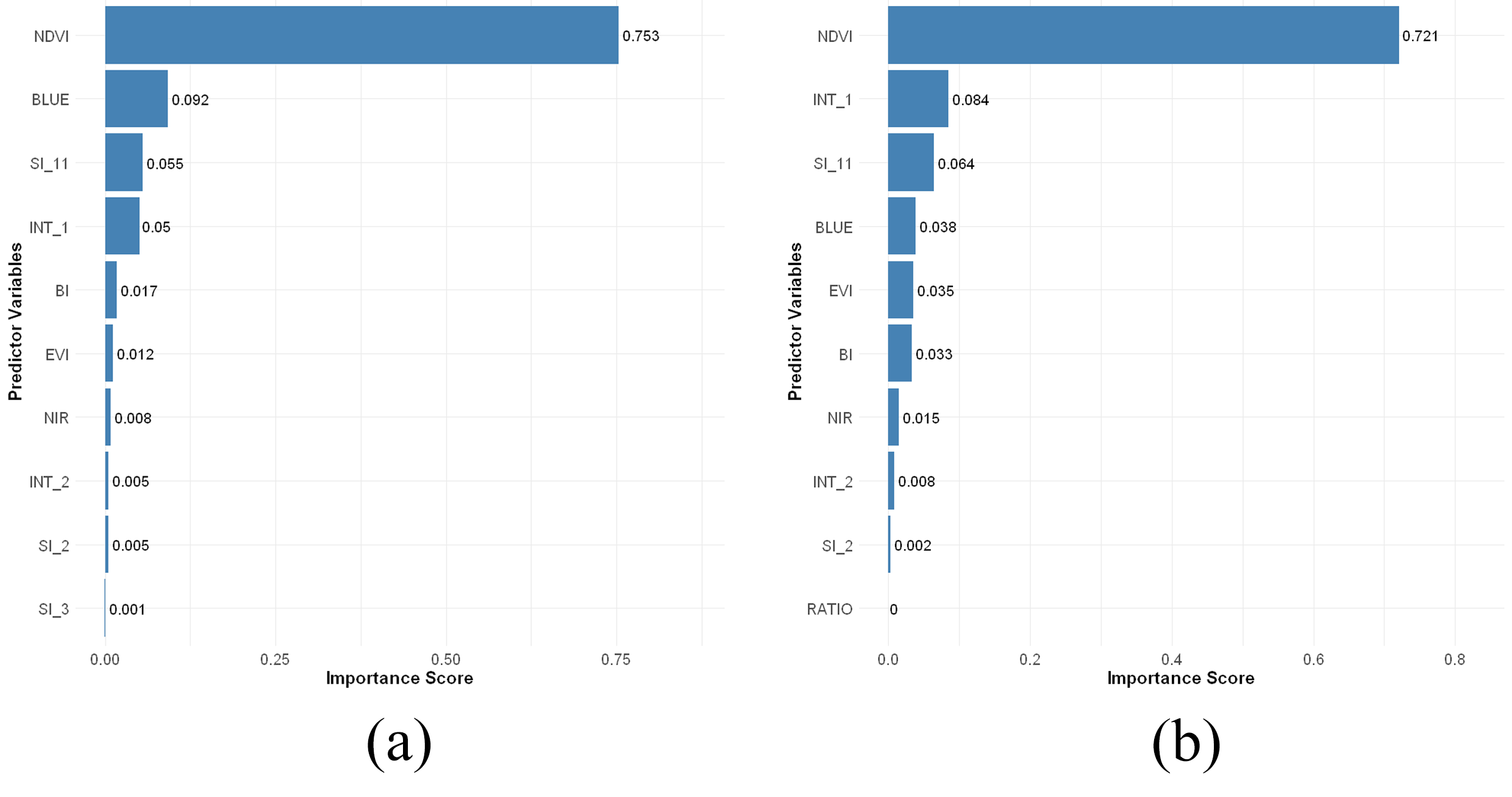}
\caption{Importance of predictor variables for soil salinity modeling in (a) 2024 and (b) 2025.}
\label{Figure 5}
\end{figure}

\begin{figure}
\centering
\includegraphics[width=\linewidth]{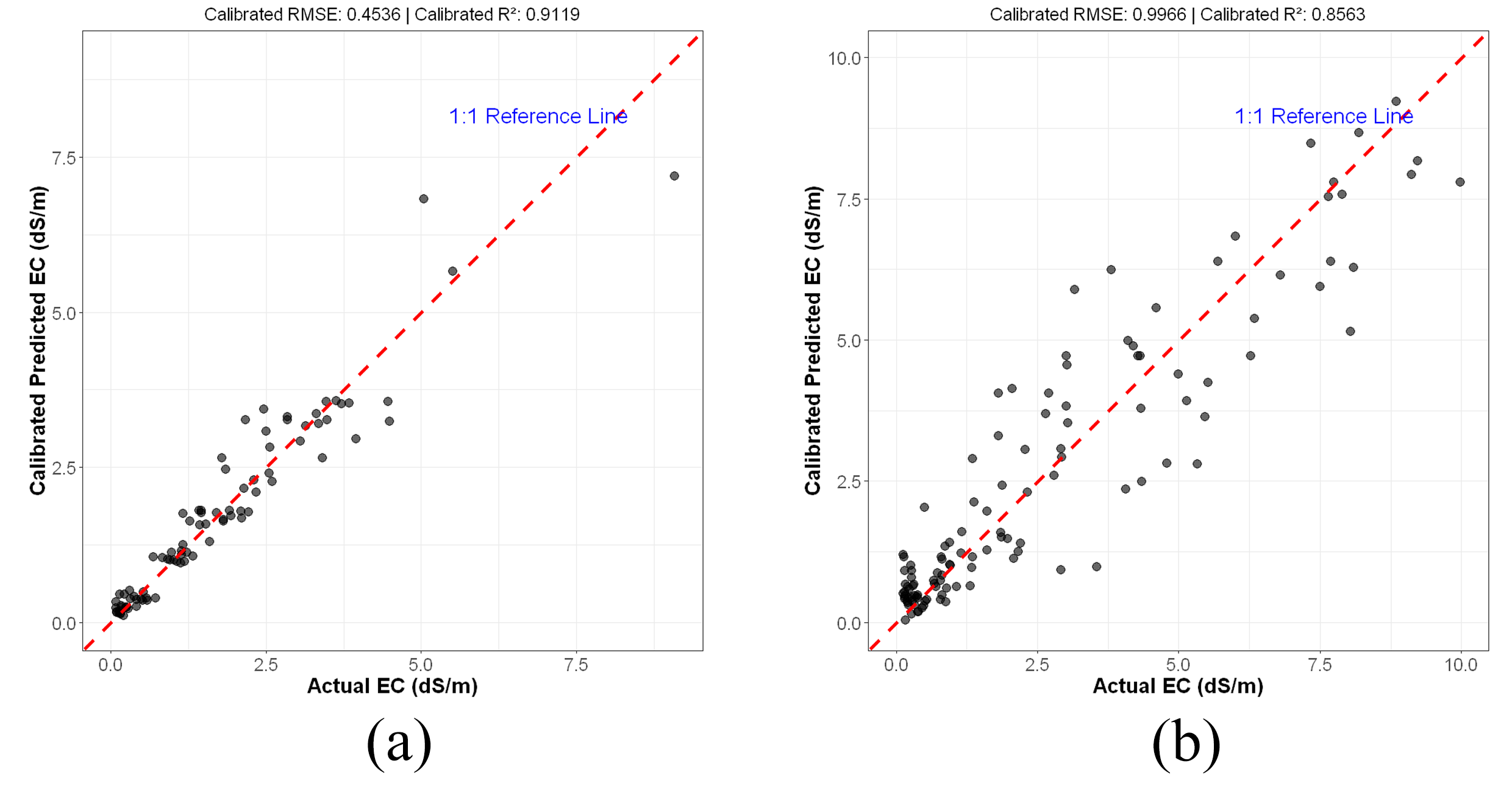}
\caption{Comparison of observed and predicted soil salinity (EC) values for model validation in (a) 2024 and (b) 2025.}
\label{Figure 6}
\end{figure}

\begin{figure}
\centering
\includegraphics[width=\linewidth]{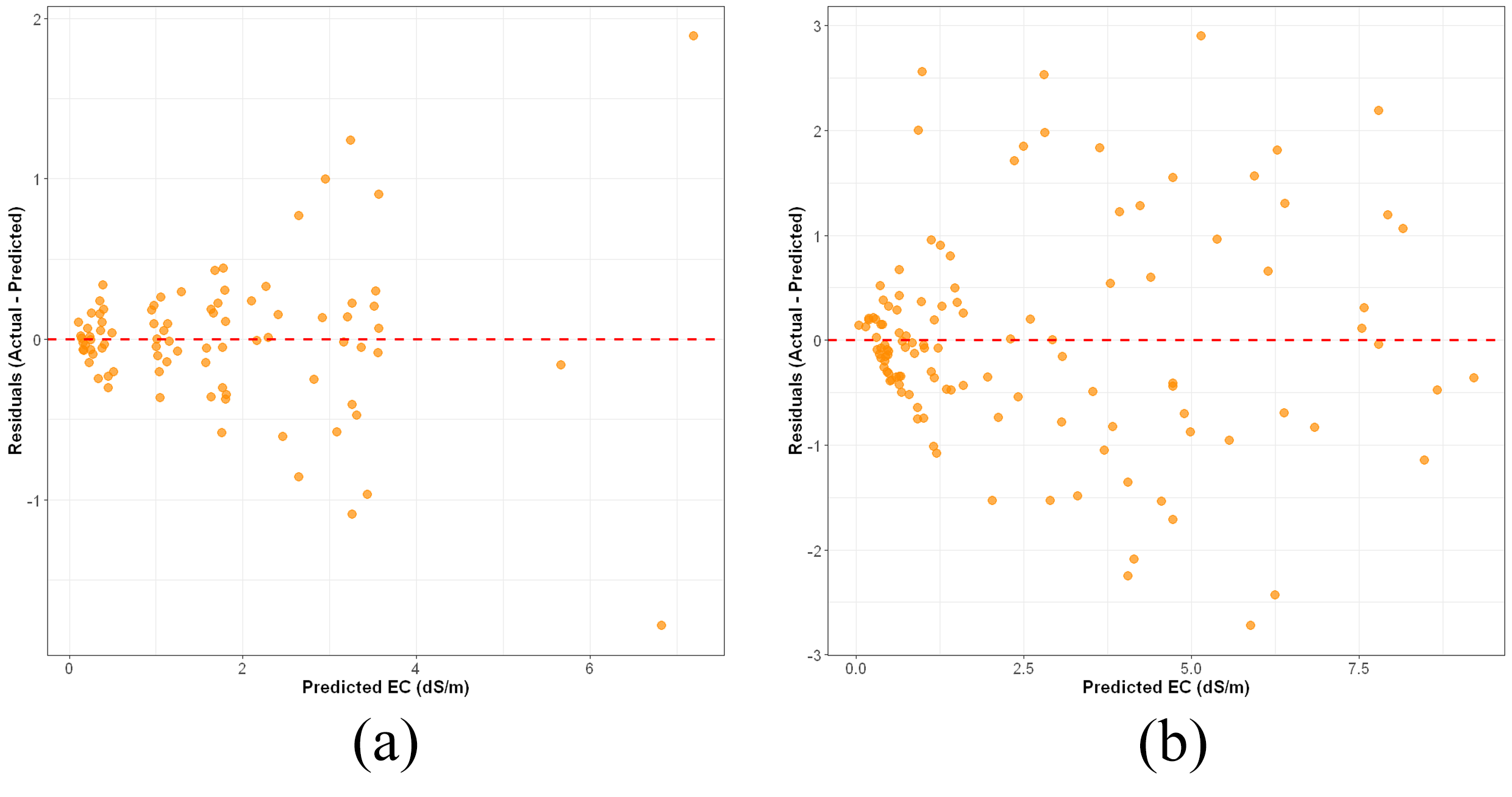}
\caption{Residual distribution of model predictions for (a) 2024 and (b) 2025}
\label{Figure 7}
\end{figure}

\begin{figure}
\centering
\includegraphics[width=5truein]{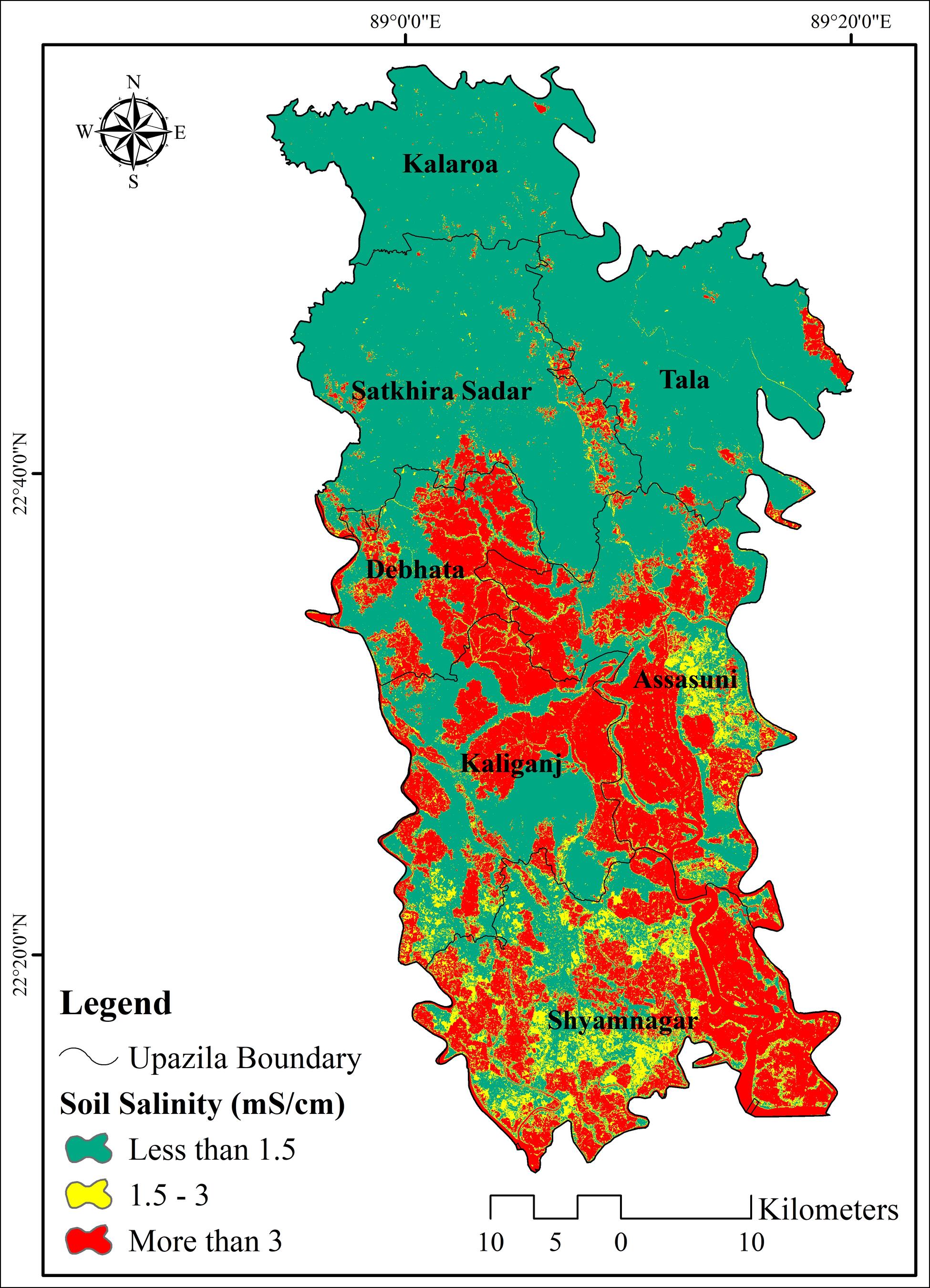}
\caption{Soil salinity peak-exposure map in 2024}
\label{Figure 8}
\end{figure}

\begin{figure}
\centering
\includegraphics[width=5truein]{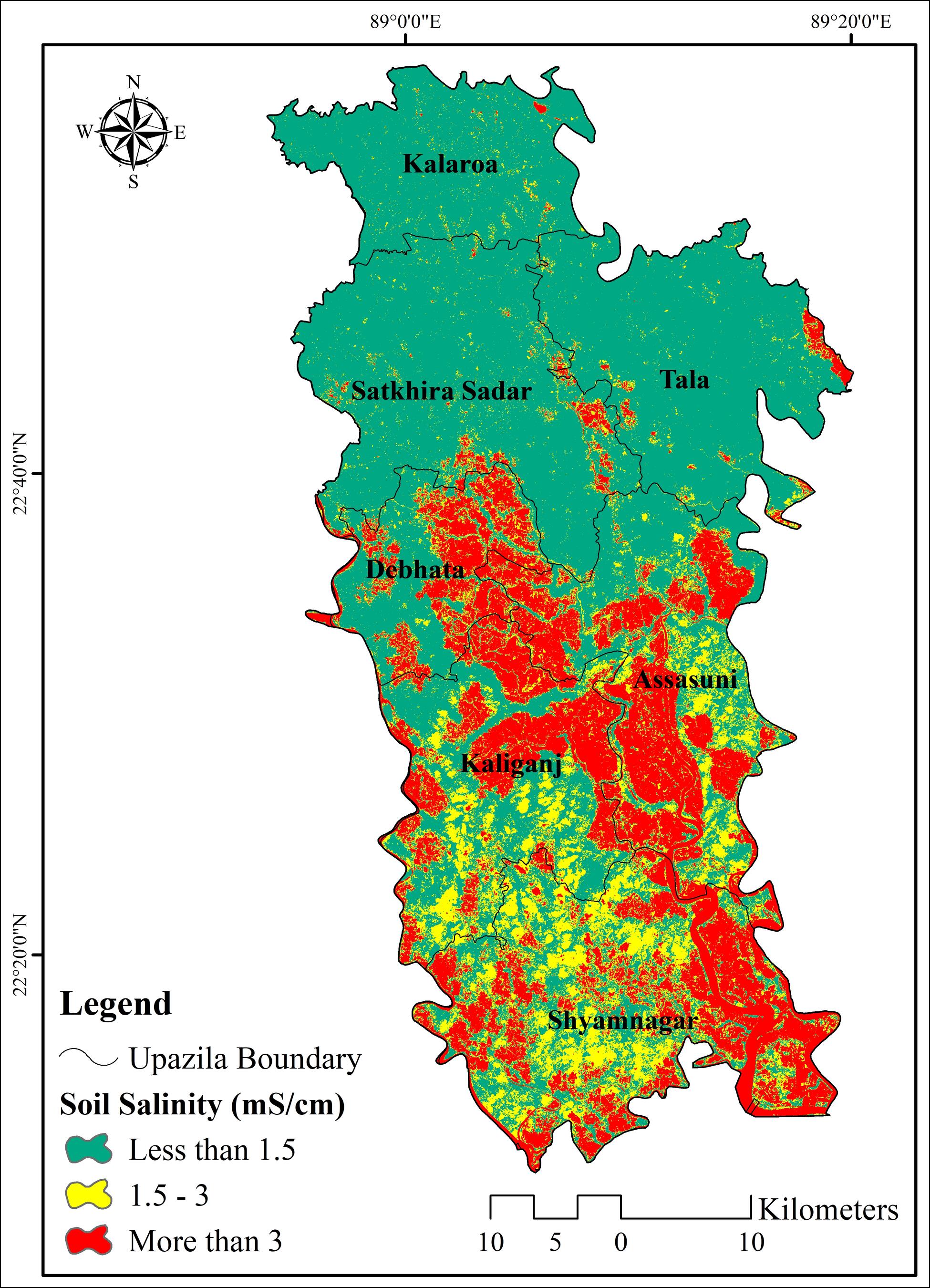}
\caption{Soil salinity peak-exposure map in 2025}
\label{Figure 9}
\end{figure}

\begin{figure}
\centering
\includegraphics[width=4.5truein]{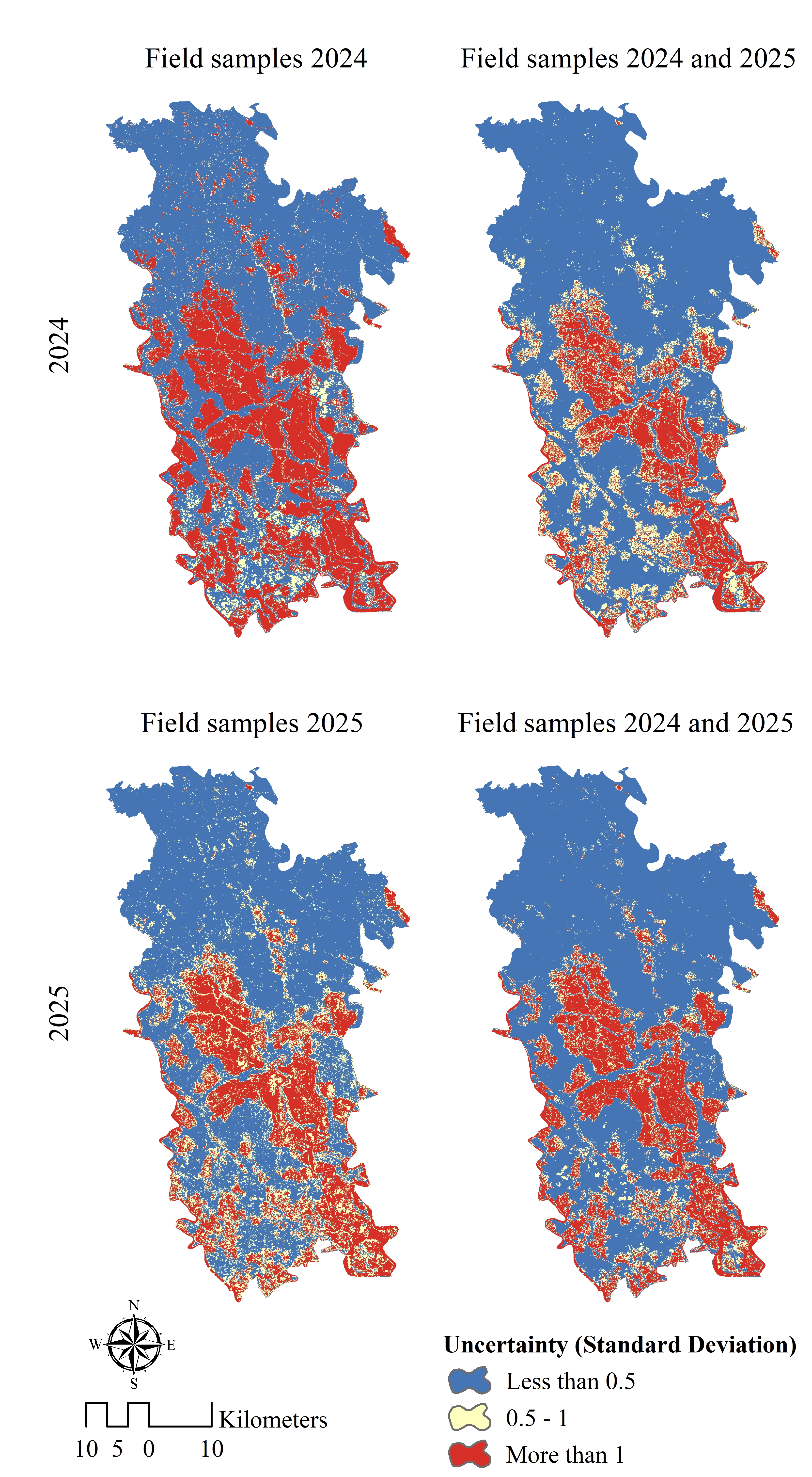}
\caption{Spatial distribution of model uncertainty for soil salinity prediction in 2024 and 2025}
\label{Figure 10}
\end{figure}

\begin{figure}[p]
\centering
\includegraphics[width=6truein]{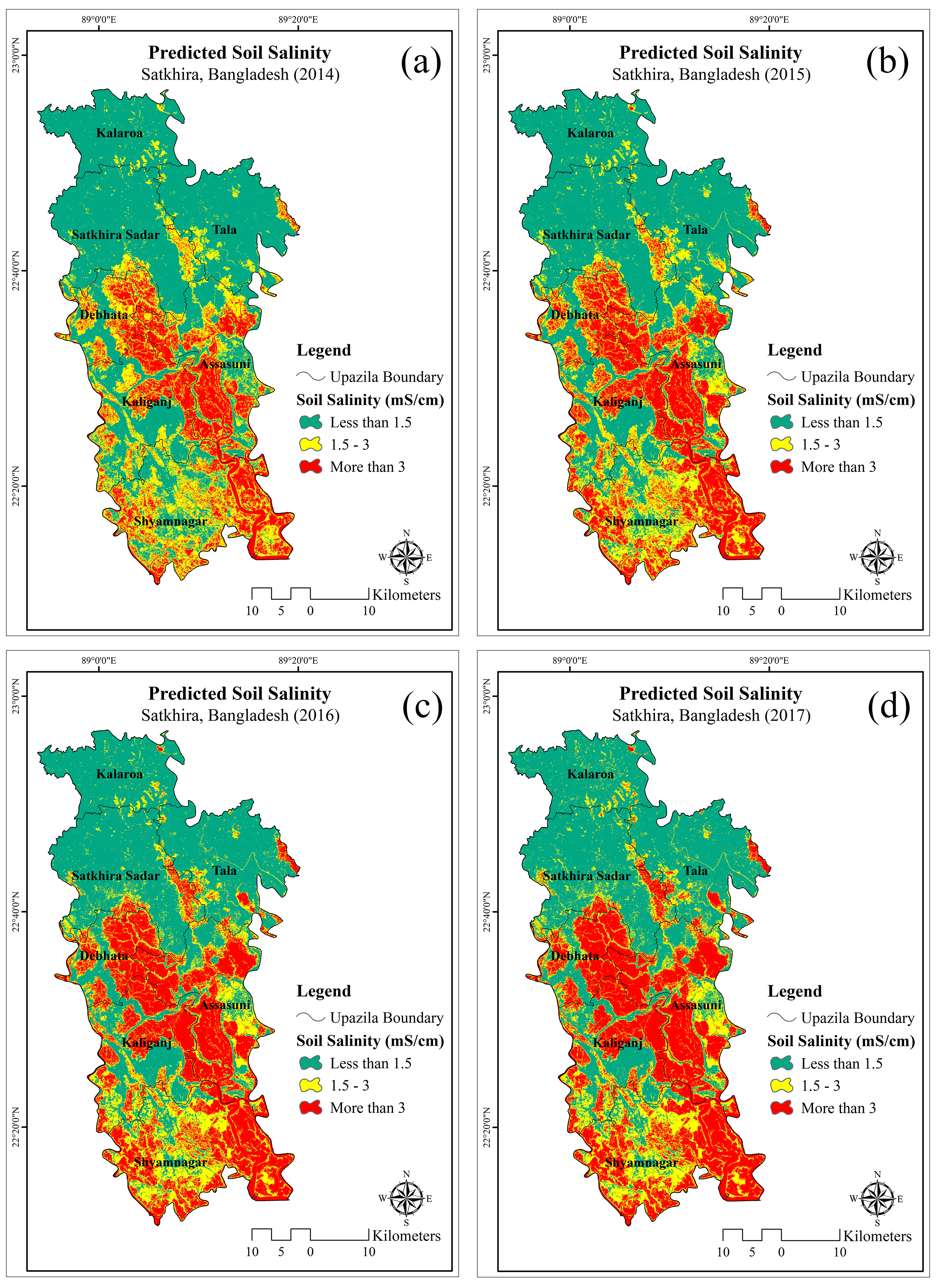}
\caption{Time-indexed peak-exposure maps of soil salinity in Satkhira district, Bangladesh for (a) 2014, (b) 2015, (c) 2016, (d) 2017, (e) 2018, (f) 2019, (g) 2020, (h) 2021, (i) 2022, and (j) 2023}
\label{Figure 11}
\end{figure}

\clearpage
\addtocounter{figure}{-1}

\begin{figure}[p]
\centering
\includegraphics[width=6truein]{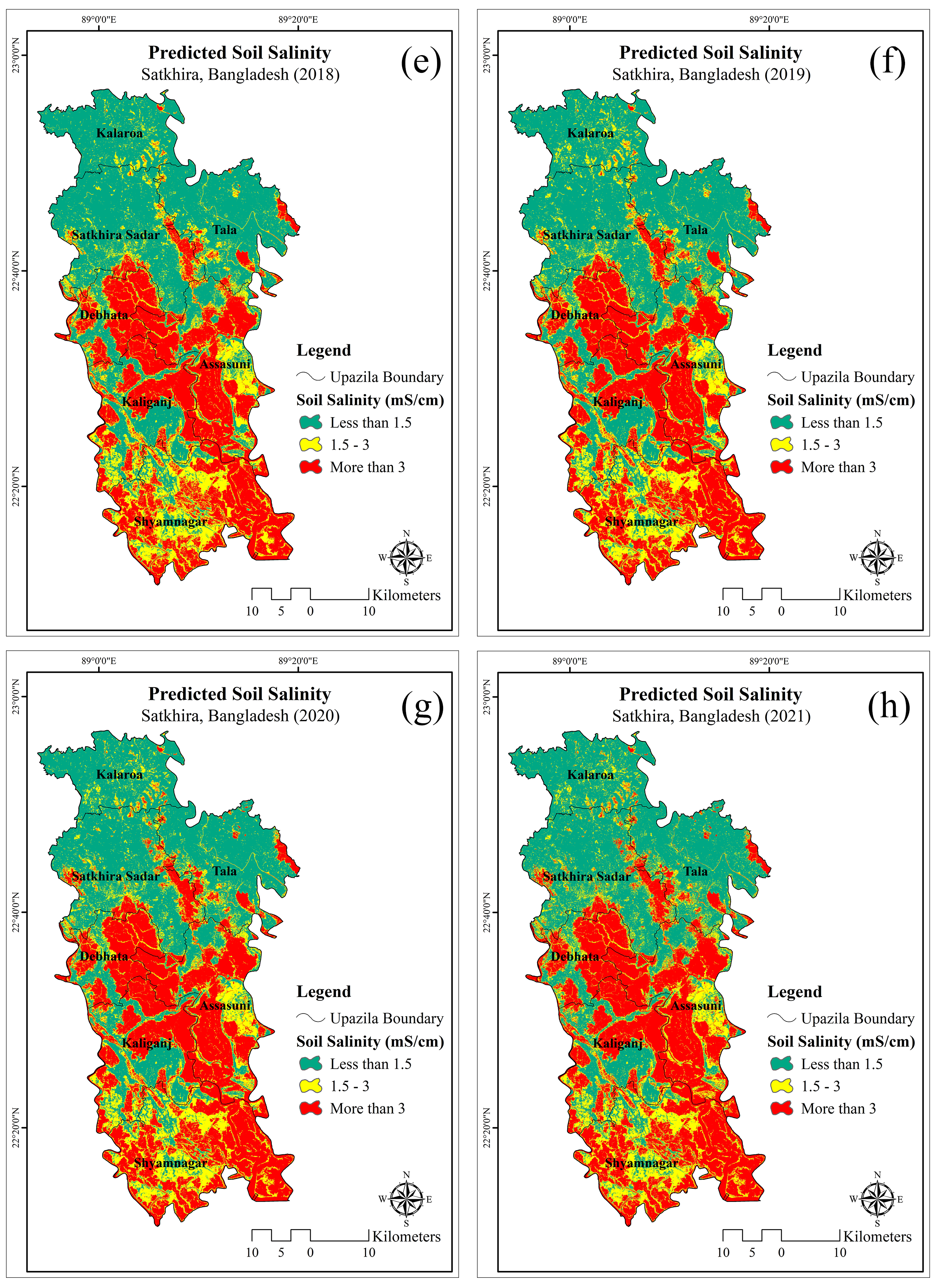}
\caption{(continued)}
\label{Figure 11}
\end{figure}

\clearpage
\addtocounter{figure}{-1}

\begin{figure}[p]
\centering
\includegraphics[width=6truein]{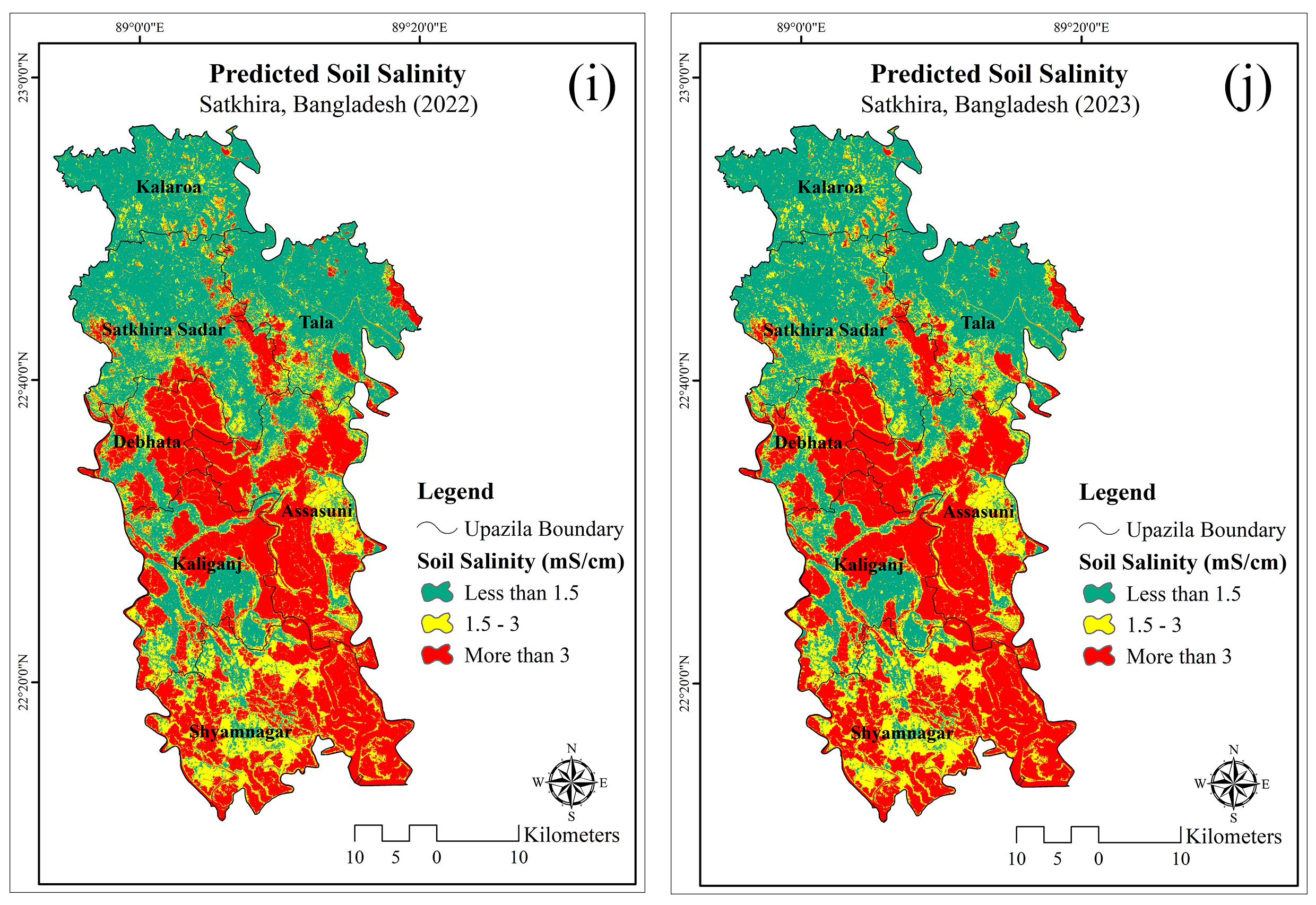}
\caption{(continued)}
\label{Figure 11}
\end{figure}

\newpage

\begin{figure}
\centering
\includegraphics[width=\textwidth]{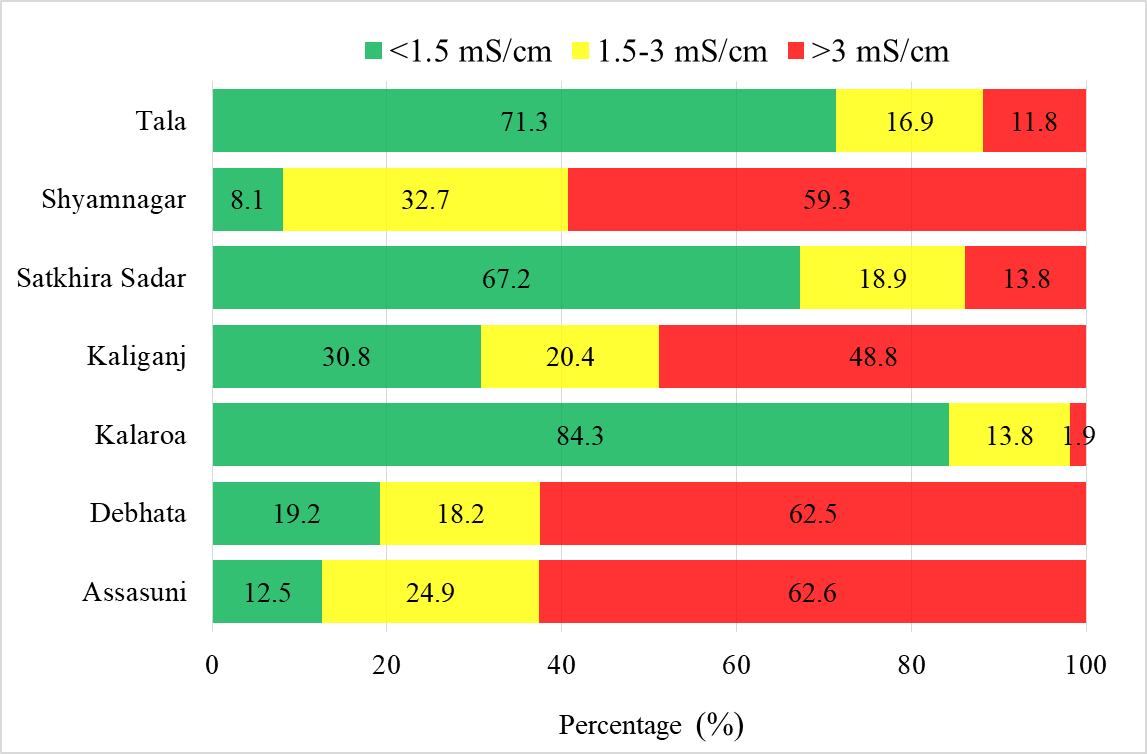}
\caption{Upazila-wise distribution of area (percent) across three soil salinity categories based on peak exposure during 2014-2023}
\label{Figure 12}
\end{figure}

\end{document}